# Real-Time Non-Invasive Imaging and Detection of Spreading Depolarizations through EEG: An Ultra-Light Explainable Deep Learning Approach

Yinzhe Wu, *Student Member*, IEEE, Sharon Jewell, Xiaodan Xing, Yang Nan, Anthony J. Strong, Guang Yang, *Senior Member*, IEEE, and Martyn G. Boutelle

***Abstract*— A core aim of neurocritical care is to prevent secondary brain injury. Spreading depolarizations (SDs) have been identified as an important independent cause of secondary brain injury. SDs are usually detected using invasive electrocorticography recorded at high sampling frequency. Recent pilot studies suggest a possible utility of scalp electrodes generated electroencephalogram (EEG) for non-invasive SD detection. However, noise and attenuation of EEG signals makes this detection task extremely challenging. Previous methods focus on detecting temporal power change of EEG over a fixed high-density map of scalp electrodes, which is not always clinically feasible. Having a specialized spectrogram as an input to the automatic SD detection model, this study is the first to transform SD identification problem from a detection task on a 1-D time-series wave to a task on a sequential 2-D rendered imaging. This study presented a novel ultra-light-weight multi-modal deep-learning network to fuse EEG spectrogram imaging and temporal power vectors to enhance SD identification accuracy over each single electrode, allowing flexible EEG map and paving the way for SD detection on ultra-low-density EEG with variable electrode positioning. Our proposed model has an ultra-fast processing speed (<0.3 sec). Compared to the conventional methods (2 hours), this is a huge advancement towards early SD detection and to facilitate instant brain injury prognosis. Seeing SDs with a new dimension – frequency on spectrograms, we demonstrated that such additional dimension could improve SD detection accuracy, providing preliminary evidence to support the hypothesis that SDs may show implicit features over the frequency profile.**

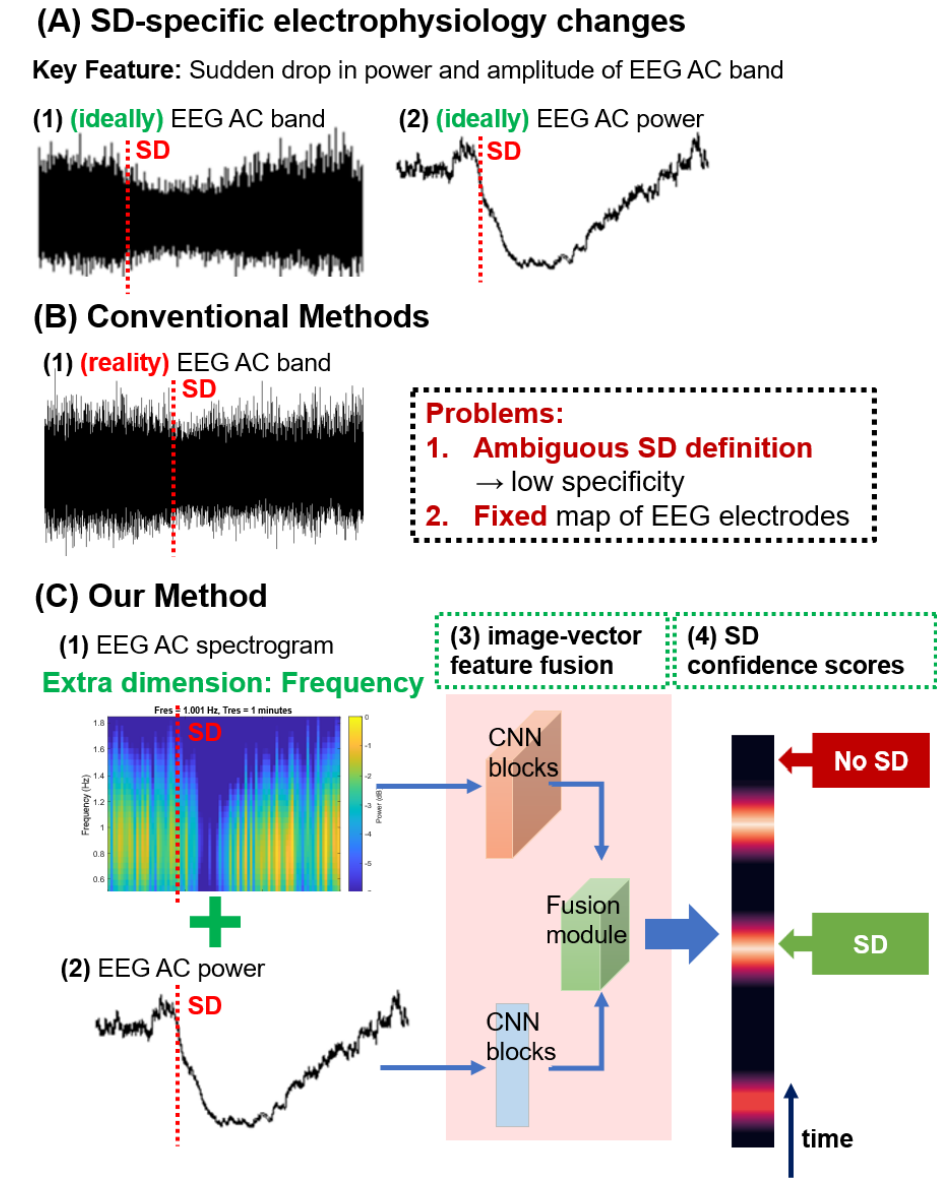

**Figure 1.** Challenges and our solutions to non-invasive detection of spreading depolarization (SD). **(A)** SD-specific features in the AC band tracing and power of EEG. **(B)** Challenges in defining SDs in EEG and limitations of the current methods [30], [33]: The conventional method which determines SD based on reduction of power temporally is prone to often fluctuating EEG AC power offsets in real-world patients, producing many false positives, where additional dimension for observation is needed. Previous methods also mandate a fixed map for network input [33] and high density of EEG scalp electrodes [34], where both are rarely possible in the clinical setting. **(C)** Our solutions: We present a novel approach to resolve the power spectrum over the frequency domain in additional to the time, allowing more features for exploitation. Our model incorporates the spectrogram imaging with conventionally used power vector for better accuracy, where we consider each node separately without any constraints on density and positioning of EEG. It also supplies confidence scores to further support clinical decision making.

## I. INTRODUCTION

ONE of the core aims of neurocritical care for severe acute brain injury patients is to minimize the occurrence of secondary brain injury [1]. Any secondary injury is likely due to expansion from the initial primary traumatic or ischemic injury region into surrounding brain tissue where blood flow is typically compromised and unreactive. Spreading depolarization (SD), was first reported by Leao et al [2] as a depression of high frequency local field potential activity that

Submission date: 23 February 2024. This work was supported in part by UK Research and Innovation (UKRI) Medical Research Council (UK) Clinical Training Fellowship (MR/R00112X/1).  (Correspondence Authors: Y. Wu and M.G. Boutelle)
Y.W., S.J., X.X., Y.N., G.Y., and M.G.B. are with Department of Bioengineering, Imperial College London, London SW7 2AZ, UK (yinzhe.wu18@imperial.ac.uk, x.xing@imperial.ac.uk, y.nan20@imperial.ac.uk, g.yang@imperial.ac.uk, m.boutelle@imperial.ac.uk). Y.W. and G.Y. are also with Royal Brompton Hospital, Sydney Street, London SW3 6LY. S.J. and A.J.S. are with Department of Basic and Clinical Neuroscience, Institute of Psychiatry, Psychology and Neuroscience, Academic Neuroscience Centre, King's College London, Room A1.27, De Crespigny Park, Box 41, London, SE5 8AF, UK (sharon.1.jewell@kcl.ac.uk, anthony.strong@kcl.ac.uk). The anonymized EEG dataset may be available upon reasonable request. Requests for data access should be directed to Y.W. at yinzhe.wu18@imperial.ac.uk

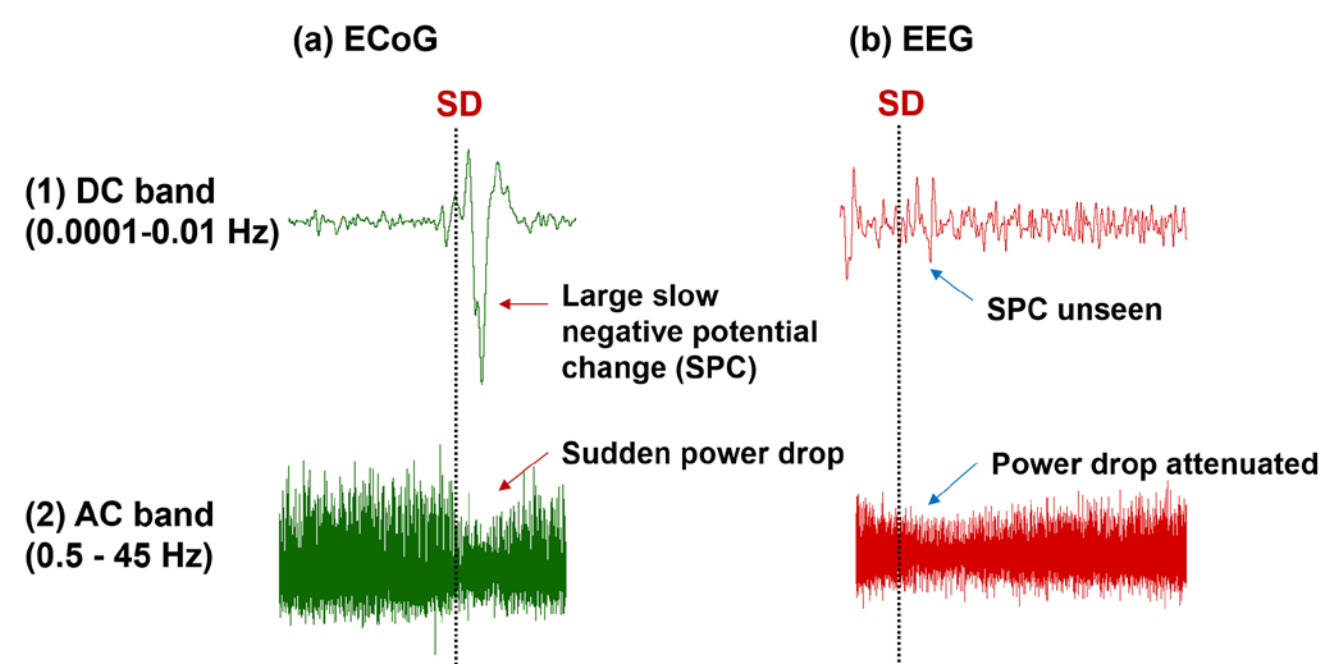

**Figure 2.** Examples of a SD occurrence in (1) near-direct current (DC)/DC (0.0001-0.01Hz) and (2) alternating current (AC) (0.5-45Hz) bands of (a) an EEG tracing and (b) an ECoG tracing for an SD occurrence. Axes: Time (Horizontal) and Power/mV (Vertical). A SD can be clearly identified from ECoG DC as a large negative slow potential change (SPC) and from ECoG AC as a drastic amplitude and power reduction. However, from the largely attenuated and noisy EEG, it remains unclear if the SPC can be reproducibly seen in EEG DC and remains challenging to differentiate the SD-specific EEG AC power reduction from other non-pathological power change.

moved slowly across the rabbit cortex, It was first reported in TBI patients [1], [3], [4] aneurysmal subarachnoid haemorrhage (aSAH) [5], malignant hemispheric stroke (MHS) [6] and intracranial haematoma (ICH) [7]. Specifically, the SD is a pathophysiological wave of mass depolarization of neurons in the cerebral grey matter [8]–[11]. It originates from neurons [12], [13] in the initial focal lesion and is characterised by abrupt, near-complete and sustained massive propagating breakdown of neuronal transmembrane ion gradients [14]. Such massive ion shifts impede the gradients to be quickly restored to its usual distributions. If failed to be restored, particularly under the poor blood flow conditions of brain injury patients, they would initiate a cascaded chain of reactions and would contribute to cerebral tissue necrosis [11], [15], [16]. The relationship between the occurrence of SDs and poor patient outcome from TBI has been clearly demonstrated and established [17], [18]. This has led SD to be considered a key driver of secondary brain injury.

The significant and diverse nature of the disruption brought by SDs [19] gives us many options to capture them [20], [21] in animal models. However, in the clinical context of neurocritical care for live patients, only observation of neural electrophysiology changes can be used routinely for reliable and reproducible bed-side monitoring of SDs. The signals of electrocorticogram (ECoG) are generated through strip electrodes placed on the cortex or through intracortical electrodes inserted through a burr hole [22], [23]. The occurrence of a SD can be seen in ECoG as **(a)** a large negative slow potential change (SPC) in the extracellular direct coupled (DC) potential **(Figure 2(a1))** and then **(b)** depressed electrophysiological activities **(Figure 2(a2))**. Although the ECoG has been superior in monitoring SD occurrences and has been considered as the 'gold standard' for the bed-side monitoring of SDs, the invasiveness of ECoG monitoring allows only a small proportion of patients with brain injury to benefit from it [16].

As the disruptive electrophysiological effects from the SDs are significant, some studies also suggested that we could see SD-induced electrophysiological changes from electroencephalogram (EEG) [24], [25], showcasing the potential of the non-invasive diagnostic tool EEG for SD screening. However, others have found that it was not possible to detect ECoG verified SD with EEG using conventional methods [26], calling for approaches using trained computational EEG analytics. EEG's advantage of non-invasiveness potentially enables intraoperative monitoring during carotid endarterectomy and SD screening for delayed cerebral ischaemia after aSAH and more importantly SD monitoring for millions of stroke patients [25]. However, in EEG, the signature negative near-DC/DC (0.0001-0.01Hz) SPCs observable in ECoG are often hard to be observed [27]–[29], as **(a)** the capacitance in the skin acts as a DC remover to attenuate SPCs and **(b)** the usually applied pre-processing techniques such as signal high pass modules and drifting remover often silence these near-DC/DC components [30] (**Figure 2(b1)**). However, the electrophysiological activities with faster frequencies in the alternating current (AC) range of 0.5-45Hz offer us another chance to monitor SDs by looking at the signal power within this frequency range, where the signature SD-induced drop in signal power is visualised in the delta band (0.5-4Hz). Nevertheless, as the brain activities captured through EEG are confounded by many other signals such as electromyography (EMG) from the muscle on the scalp and are attenuated by the skull, it may be worth remarking that the AC EEG is much more chaotic, and thus it is very hard to find SDs from EEG AC reliably by traditional methods such as thresholding. In addition, it is concerned that the much noisier EEG may not perform well in excluding SDs for negative control subjects. Also, the poor signal quality of EEG leads to concerns over its accuracy in SD detections particularly for TBI patients where EEG electrodes could be insulated by the subcutaneous air introduced by the original or (more likely) surgical trauma.

### *A. Our Contribution*

Currently available methods only employ the 1-D time series of EEG for either identification of DC changes for SPC or signal power changes, where their limited success is prone to the incorrect assumption that EEG AC power offset is always constant (**Figure 1B**). It may worth noting that, for brain trauma patients, their frequency axis may exhibit specific changes at the time of SDs [29], [31], where additional considerations of power changes resolved over the frequency axis may offer us a new prospective for SD identification, particularly when the SPC or power changes are not visually significant over noise (**Figure 2**). This will give us a new modality – spectrogram imaging for SD detection. To consider the conventional 1D time series feature along with our proposed 2D spectrogram imaging, we employed an image-vector fusion module to read features from both spectrogram imaging and EEG power vector (**Figure 1C**).

To demonstrate the detection model's performance for more generalised use, where the model may see EEG AC power offset alternating, based on real-world EEG data collected from patients, we simulated a wide range of SD-carrying EEG tracings where it contains different levels of noise and SD characteristic AC power suppressions and offsets.

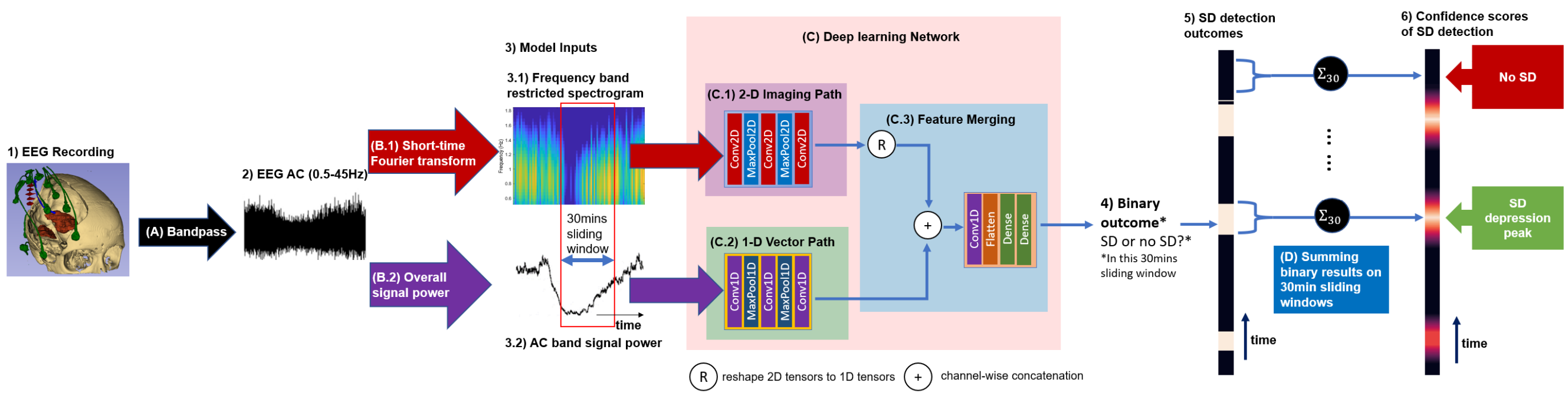

**Figure 3.** The general framework for identification of SD over an EEG tracing from a single electrode.
**(Stage 1)** The EEG of a scalp electrode first **(A)** undergoes bandpass to obtain **(Stage 2)** its AC segment and then calculated for **(Stage 3)** its **(3.1)** spectrogram images and **(3.2)** signal power vectors by **(B.1)** short-time Fourier transform and **(B.2)** overall power calculations in AC band respectively. These spectrogram images and signal power vectors were partitioned by 30-minute sliding windows. The partitioned images and power vectors were input into the **(C)** deep learning convolutional neural networks **(C.1)** 2-D image path and **(C.2)** 1-D vector path respectively before merged through **(C.3)** a feature merging block. The model would then produce **(Stage 5)** a series of binary outcome on whether there is any SD 30 minutes around each time point or not. This series of binary outcome would then be **(D)** summed through a 30-min sliding window to get **(Stage 6)** the confidence scores for the SD detection results and highlight the time position of the SD depression peak.

Previously proposed methods often assume a fixed map of multiple electrodes. However, this is not often clinically feasible. Particularly, for ICU trauma post-surgery cases, it would be extremely difficult to fix multiple electrodes on the operated patient's scalp reliably, and obviously a fixed map of EEG electrodes cannot be assumed for all patient populations [28], [32]. This calls for a method allowing more versatile electrode setup. To expand the generalisability of the detection method and allow more flexibility in the number of electrodes over the EEG map, instead of assuming a fixed EEG map, this paper focuses on accurate detection of the SD over a single EEG electrode.

We then built a novel network **(Figure 3)** to combine the 2-D spectrogram imaging input with the 1-D EEG AC power wave input to enhance the performance of SD identification and demonstrated its success on our datasets. Through further ablation tests, where we tried to input the model with only EEG spectrogram or the 1-D EEG AC power wave, we found that the model performs better when input with the spectrogram compared to the power wave. This demonstrated success of EEG spectrogram as a stand-alone independent biomarker for SD, which could offer better SD detection performance from its additional axis in frequency.

To enable further expanded clinical use and benefits, real-time detection of SD is the key for early SD detection and prognosis of secondary brain injury. However, all previously proposed algorithm failed to demonstrate its ability for real-time detection of SD under standard clinical setting where no GPU is available [30], [33], [34]. The conventional methods proposed by Chamanzar et al. required 2 hours for processing over CPU [30], [33], and their proposed deep learning methods required GPU for processing [34], where GPU is not always available in bed-side machines for expanded clinical use. In this study, our ultra-light deep learning model is specifically designed to enable itself to be continuously inferred over a CPU with a dramatically reduce processing time (<0.3 sec per each hour of EEG), which is a huge advancement for expanded clinical use and translation compared to the processing time of the conventional methods.

Conventional methods need 30-hour EEG [30], [33] for decisions, not feasible clinically. Our model uses 30-minute window, reducing pressure for monitoring and aiding patients

Existing deep learning methods for SD lack transparency. By linking confidence scores with outputs, our model enhances model trustworthiness and explainability and clinical decision-making.

The key merits of this paper are that we

**(1)** simulated EEG tracings for SD patients carrying altering AC power offsets from clinical cases,

**(2)** built a novel ultra-light-weight deep learning model fusing features in both EEG spectrogram and EEG power wave for more accurate SD detection with confidence scores for more explainable decisions,

**(3)** dramatically reduced the processing time (**<0.3 sec** per each hour of EEG) by the model's ultra-light weight and became the first real-time SD monitoring framework on CPU

**(4)** dramatically reduced time window (30 minutes vs. (conventional) 30 hour) for final decisions, enabling quicker delivery of decision, reduced bed-side monitoring pressure and easier and expanded use, and,

**(5)** through the deep learning network's performance, demonstrated the efficacy of spectrogram as an independent imaging biomarker for more accurate SD identification.

As the method reported in this paper allows detection of SDs through individual scalp electrodes with no limitation of their amounts or positions (**Figure 1C**), we consider our paper to be the first step towards detecting SDs over ultra-low-density EEGs of variable positions, allowing more flexibility and much lower density in the clinical installation of EEG for non-invasive SD detection. More importantly, the ultra-short processing time and the light weight of the model framework makes it the first real-time non-invasive SD detection framework for expanded clinical use and translation.

## II. Related Work

EEG is a critical tool in diagnosing neurological diseases [35]. Due to the scarcity of human experts and the increasing demand for EEG interpretation [36]–[38], there has been a significant push for automated EEG analysis [39]–[41]. This automation aims to broaden and accelerate access to EEG's benefits, where multiple conventional strategies such as wavelet analysis, machine learning methods such as support vector machine, and advanced deep learning methods such as convolutional neural networks have been widely adopted [42], [43]. Specifically, in the realm of electrophysiology and disorders like epilepsy, machine learning and deep learning models have shown considerable promise. They have been extensively researched for various applications, such as differentiating between normal and abnormal EEG recordings [44], detecting seizures [45]–[47], and identifying interictal epileptiform discharges (IEDs) [48]. These models have been particularly effective in detecting IEDs [49]–[52], with publicly available data aiding in their training and validation [53]. A recent publication has further expanded the scope of EEG analysis in epilepsy detection [54], showcasing a comprehensive approach validated across a broader patient population. This advancement marks a significant step in the field, enhancing the robustness and applicability of EEG-based diagnostic methods for epilepsy.

However, it's important to note that spreading depolarization (SD) is **distinct** from these neurological diseases commonly addressed by these automated detection methods, in terms of their drastically different electrophysiological event timescales and features [55]–[57]. This prevents the wide success of machine learning and deep learning model developed for other diseases to be directly applicable, transferrable to, or benchmarked in non-invasive SD detection in EEG.

In the specific context of non-invasive SD detection in EEG, there is no publicly available dataset and the reliable concurrent acquisition of EEG and ECoG from TBI patients are particularly difficult. Recent studies showed limited success using unrealistic assumptions, through adaptive thresholding via power envelopes [30], [33] or a ResNet-graphical neural network combo [34] on EEG power traces. This lack of extensive research in the field, rather than being a drawback, actually highlights the novelty and significance of our article's contribution in such emerging neurology field. Chamanzar et al [30], [34] achieved SD detection in simulations, but their fixed high-density EEG setup (40 electrodes per patient) isn't always feasible. Particularly, for ICU trauma post-surgery cases, it would be extremely difficult to fix multiple electrodes on the operated patient's scalp reliably, and obviously a fixed map of EEG electrodes cannot be assumed for all patient populations [28], [32], which calls for a method allowing analysis per each single electrode without restricting electrode setup. Their method relies on constant EEG AC power offset [30] and SD signature power suppression, which clinical cases don't always follow. Despite validation on a small patient group [33], their power envelope-optical flow method overlooked broader clinical complexities, being sensitive to EEG amplitude outliers and causing many false positives (**Figure 1B**).

For prompt SD detection and injury prognosis, real-time detection is vital. Past algorithms failed to deliver real-time results in standard clinical setups without GPUs [30], [33], [34]. Their power-based and optical flow approach takes 2 hours per CPU inference, delaying reporting. Use of GPU was able to accelerate the processing time to 5 minutes [33], but remained costly and not widely available for bedside clinical use.

Moreover, their power envelope and optical flow approach [33] needs a 30-hour EEG window for decisions. Yet, gathering such extended, uninterrupted EEG of high quality for the model's inputs is impractical in real-world clinical settings.

## III. Methods

To address the need for more adaptive SD detection framework with the extra observation dimension and the primary goal for early SD detection and to facilitate instant brain injury prognosis, we developed the general framework for generation of the frequency band restricted spectrograms and identification of SD over an EEG tracing from a single electrode as described in this section (**Figure 3**).

### *A. Generation of Frequency band restricted EEG Spectrogram and Temporal Power tracing*

Currently available methods only employ the 1-D time series of EEG for either identification of DC changes for SPC or signal power changes, where their limited success is prone to the incorrect assumption that EEG AC power offset is always constant (**Figure 1B**). It may worth notice that, for brain trauma patients, their frequency axis may exhibit specific changes at the time of SDs [29], [31], where additional considerations of power changes resolved over the frequency axis may offer us a new prospective for SD identification, particularly when the SPC or power changes are not visually significant over noises (**Figure 2**).

The key impression of SDs in AC (0.5-45Hz) range is its sudden neural electrophysiological silence, which could be characterized by the signal power over time. To further exploit its implicit features over different frequencies, we resolve the EEG signal by short-time Fourier transform (STFT) to form spectrograms.

STFT is a time-frequency analysis tool for non-stationery signal analysis, which partitions signals into temporal segments by windowing and apply Fourier transform on these. STFT of a signal $s(t)$ can be denoted as $\mathcal{F}(\tau,\omega)$, as per Equation (1).

$$\mathcal{F}(\tau,\omega)=\int_{-\infty}^{\infty}s(t)h(t-\tau)e^{-j\omega t}dt \quad (1)$$

, where $h(t)$ is the windowing function. The power spectral density over the spectrogram can then be defined as per Equation (2), which represents the power of EEG resolved at different time $\tau$ and frequency $\omega$.

$$P_s(\tau,\omega)=|\mathcal{F}(\tau,\omega)|^2 \quad (2)$$

For this study, Hanning window was used in STFT for spectrogram rendering, with the time resolution set as 1 minute (to alleviate computing pressure in subsequent models while maintaining its latent information as SDs are usually much longer than that [23], [30], [34]) and at a frequency resolution of 1 Hz (to allow sufficient differentiation between frequency bands). (**Figure 4A**). Upon formation of the spectrogram, we then examined the signal power at different frequency bands.

Confirmed by the persistence spectrum of the signal (**Figure 4B**), which shows the distribution of signal power at different frequencies, the EEG signals are much stronger in the delta frequency band (0.5-10Hz), which is a common feature of the brain injury [25]. This can be explained by the preserved but abnormal synaptic activity due to structural damage or reduction of cerebral blood flow [25]. Particularly, our data has its highest power around 1 Hz (**Figure 4B**). Thus, we extracted specifically the power arrays for the frequencies between 0.5 and 1.85 Hz for evaluation of the frequency band carrying the greatest power spectrum with the greatest power changes while entering the SD.

The generation of the temporal EEG power follow the same windowing function as STFT, where the total power within the AC frequency range in each time window is calculated.

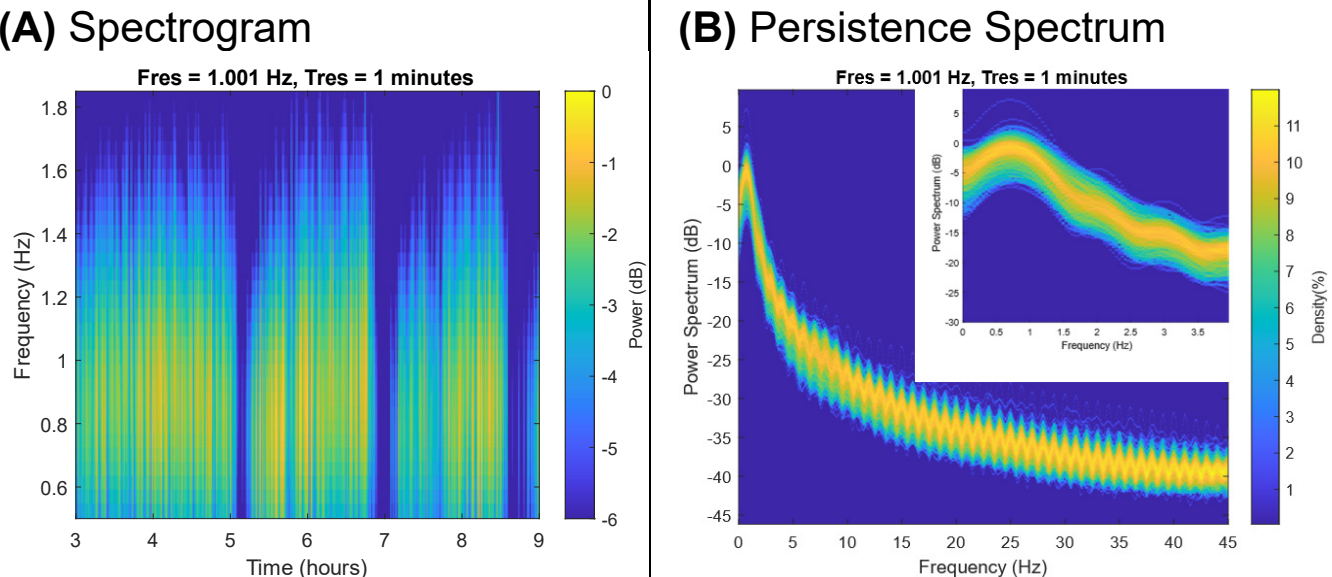

**Figure 4. (A)** Spectrogram of SD occurrences, where the power suppression could be seen clearly on the spectrogram (near 5th, 7th, 9th hour), and where a differential frequency profile could be observed during the SD-specific suppression and recovery of signal power. **(B)** Persistence Spectrum of a typical SD carrying EEG recording (Upper right corner: a zoomed in view for frequency band [0, 4] Hz). We could observe the signal power is the highest around 1 Hz. For both **(A)** the spectrogram and **(B)** the persistence spectrum: frequency limits: [0, 45] Hz; frequency resolution: 1 Hz; time resolution: 1 min

## *B. Windowing of the spectrogram and power vectors for model analysis*

For input into the model, we crop the spectrograms and the power vectors by a 30-min moving window for input into the model in **Stage 3 (Figure 3.3)**, as it could usually cover most of each SD episode [32], so the model can consider if there is a SD within the given 30-min time window around each time point. Upon cropping, the frequency restricted spectrogram images become a series of 30 × 30 image arrays, and the power vectors become a series of 1-D arrays of size 30

## *C. Network architecture and settings of the ultra-light-weight SD detection model*

Upon generation and normalization of the spectrogram images and temporal power vectors, they got input into a series of convolutional neural networks (CNN) (**Figure 3C**).

The clinical need for real-time online processing over bed-side clinical machines mandates the light weight of the model and instant processing of results without high-end hardware such as GPU, which is not generally available in bed-side monitoring devices. Although Transformer or attention-based models may boost the model detection performance, large complex model is not appropriate for use in this clinical scenario, where they require GPU over the bed to process all these high frequency sampled EEG. For our developed model, the average inference time per each hour of EEG is 4 msec if with a GPU or 0.13 sec if with a CPU only, which fulfils the requirement for real-time bed-side processing. Even if adding the preprocessing time (0.08 sec per each hour of EEG in average), the total processing time is still less than 0.3 sec in average per each hour of EEG if with a CPU only. Further discussions of the model inference time are elaborated in **IV.D.3)**.

Inspired by a recent success in image-text embedding [58], we adopt similar network architecture for initial image and vector processing. The spectrogram images are input into 3 blocks of 2D convolution layers with a kernel size of 3×3, a stride of 1×1 and the ReLU [59] activation function and a Max Pooling Layer with a stride of 2×2 (**Figure 3C.1**). Similarly, the power vectors are input into 3 blocks of 1D convolution layers with a kernel size of 3, a stride of 1 and the ReLU [59] activation function and a Max Pooling Layer with a stride of 2 (**Figure 3C.2**).

To merge the outcome of the image 2-D CNN chain and the vector 1-D CNN chain, the image 2-D CNN chain got reshaped into 1-D before it is concatenated with the final CNN of the vector CNN processing chain (**Figure 3C.3**). A further CNN is attached after concatenation for fusing features in spectrogram imaging and power vectors (**Figure 3C.3**). A summary of tensors in this CNN chain can be found in **Figure 5**.

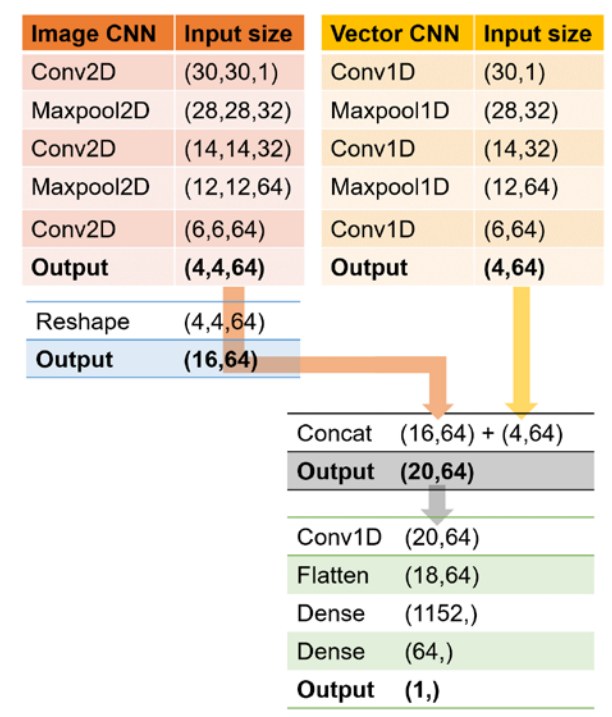

| Image CNN | Input size |
|---|---|
| Conv2D | (30,30,1) |
| Maxpool2D | (28,28,32) |
| Conv2D | (14,14,32) |
| Maxpool2D | (12,12,64) |
| Conv2D | (6,6,64) |
| **Output** | **(4,4,64)** |

| Vector CNN | Input size |
|---|---|
| Conv1D | (30,1) |
| Maxpool1D | (28,32) |
| Conv1D | (14,32) |
| Maxpool1D | (12,64) |
| Conv1D | (6,64) |
| **Output** | **(4,64)** |
| Reshape | (4,4,64) |
| **Output** | **(16,64)** |
| Concat | (16,64) + (4,64) |
| **Output** | **(20,64)** |
| Conv1D | (20,64) |
| Flatten | (18,64) |
| Dense | (1152,) |
| Dense | (64,) |
| **Output** | **(1,)** |

**Figure 5.** Parameter details of the CNN detailed in Figure 3(C)

A binary outcome at **Stage 5** is produced after the CNN fusion layer (**Figure 3.4 and 3.5**) and then analyzed through the method detailed in **III.E**.

## *D. Restoration of windowed outcomes and generation of confidence scores*

It is often complained by the clinicians that they often fail to understand the rationale of decisions made by deep learning models. To support such need for clinical decision making further, in line with [60], we supply quantitative confidence scores in addition to the qualitative detection outcomes, which all previous methods fail to provide.

As explained in **III.C**, each binary outcome refers to the existence of SD within 30min around that time point. Accordingly, for a SD peak, we would expect positive binary outcomes within the 30-min time frame around that peak, where a sliding summing window could help to highlight the position of the peak. Upon gathering all detection outcomes in a time series in **Stage 5 (Figure 3.5)**, we use another 30-min moving window to add up the binary outcomes around each time point **(Figure 3D)**. This will also create a 0-30 score at **Stage 6 (Figure 3.6)** to give an impression of the model's confidence of its SD recognition at a given time point on the EEG.

## IV. Patient Studies

### A. Datasets and Experiment Details

#### 1) Patient Recruitment and Data Collection

Two patients were enrolled at King's College Hospital (London, UK). Inclusion criteria were the clinical decision for the neurosurgical craniotomy and age 18-80 years. Patients with Glasgow Coma Scores (GCS) under 4 with bilateral fixed and dilated pupils are excluded, as research monitoring in such patients is ethically inappropriate. Patients were all in pharmacologically induced coma after sedation for ventilation and monitoring [63]. As patients were all in coma at the time of admission, written assent for study participation was sought from legally authorized representatives. Consents were subsequently acquired from the patients once their mental capacity was restored when followed up. This study was approved by the KCH Research Ethics Committee (Cambridgeshire South; 05/MRE05/7) and the UC Institutional Review Board (2016-8153). The research was conducted in line with the Declaration of Helsinki.

As we need to have both EEG and ECoG connected to eligible patients, this added extra difficulty for the data collection. Although we attempted to collect EEG data from 9 patients, we only succeeded in acquiring EEG and ECoG data simultaneously from 2 patients, whereas the remaining patient population either had air injected subcutaneously or electrodes detaching during EEG monitoring, preventing reliable data collection. The patient demographics are described as below:

Patient #1 is a 28-year-old female who had a road traffic accident and suffered TBI, who had been confirmed to have 17 SDs while being monitored by ECoG. Patient #1 had been recorded for ECoG and EEG for 39 hours. Patient #1 recovered well with an extended Glasgow Outcome Scale of 6. Patient #2 is a 48-year-old male who had an MHS, who had been confirmed not to have any SDs while being monitored by ECoG. Patient #1 had been recorded for its ECoG and EEG for 22 hours. Patient #2 died. Although two subjects had different types of injuries, they had the same operation of craniotomy with their skull removed while being monitored.

When concluding the surgery, a subdural electrode strip consisting of 6 platinum electrodes (10mm centre-to-centre, Adtech, Racine, WI, USA) was placed on the surface of the cortex. Subsequent to the conclusion of the surgery, EEG electrodes were placed as a dense array over the lesion region. Following the completion of the surgery, the probes were connected to monitoring in the intensive care unit (ICU) where data acquisition started. ECoG data was acquired through the amplifier Neuralink [64] (USA). ECoG and EEG data were recorded with LabChart software (ADInstrument, Sydney, Australia). At the of monitoring, both strips and depth electrodes were removed at the bedside using gentle traction. No complication was found to be associated with the placement or removal of the electrodes.

We took signals from 2 ECoG electrodes and 4 EEG electrodes of Patient #1 and signals from 2 ECoG electrodes and 2 EEG electrodes of Patient #2 for the analysis below. Upon removing the EEG/ECoG segments with significant sudden drifting and artefacts, we have obtained a SD carrying 4-channel EEG recording along with a 39-hour 6-channel ECoG from a TBI case (Patient #1). In Patient #1, we could label 12 SD occurrence found in EEG with reference to the high resolution ECoG recording, as confirmed by the clinical electrophysiologist. We also have a 17-hour 12-channel EEG recording excluded from SD from a case with MHS (Patient #2), as confirmed by the clinical electrophysiologist.

#### 2) Pre-processing

The signature negative near-DC/DC (0.0001-0.01Hz) SPCs observable in ECoG are often hard to be observed [27]–[29], as **(a)** the capacitance in the skin acts as a DC remover to attenuate SPCs and **(b)** the usually applied pre-processing techniques such as signal high pass modules and drifting remover often silence these near-DC/DC components [30] (**Figure 1(b)**). We then only considered the AC frequency band (0.5-45Hz) for analysis, where frequencies higher than 45Hz are removed as well since it is close to the 220 Volt power supply's 50 Hz frequency. Thus, we imposed a bandpass for all signals with its lower limit as 0.5Hz and its higher limit as 45Hz. Its high pass modules also remove the drifting of the EEG signal. We then removed any abnormally high spikes over the EEG tracings, which is usually considered as signal artefacts. The signal is then normalized before further processing to ensure the generalizability of the method.

#### 3) SD detection performance over augmented SD carrying EEGs

As discussed in **IV.A.1)** it remains difficult to obtain co-registered EEG and ECoG data from further patients, for the purpose of this paper, we need to generate further data through augmenting from these 2 EEG recordings to demonstrate the performance of our model. We took 6 SD carrying segments for augmenting the data for the training purposes and 6 SD carrying segments for augmenting the data for the testing purposes. Each generates 1,700 hours (containing 918 SDs) of EEG training data and 1,700 hours (containing 951 SDs) of EEG testing data. The exact strategy to augment these data is as detailed below.

The simulated EEG data presented in previous studies either refer to the mathematical model of SD or is based on full band signal power suppression [30], [34]. Either of these two approaches failed to mimic the true SD profile over the frequency domain and is unlikely to truly reflect the fluctuating background noise of EEG. To generate SD carrying EEG series, we used the non-SD EEG tracing as the basis and add the SD signature AC power drop profile onto it by multiplying the SD carrying EEG segment onto the non-SD EEG in a weighted way with noises added on.

For each segment of EEG to be augmented by the SD specific power suppression, the simulated SD carrying EEG segment can be denoted as $\hat{s}_s$ as detailed in Equation (3).

$$\hat{s}_s = BPF_{0.5-45Hz}\left(\frac{1}{1+\beta}\left(\frac{\hat{s}_{noSD} + \alpha\hat{s}_{noSD}|\hat{s}_{SD}|}{1+\alpha} + \beta\hat{n}\right)\right) \quad (3)$$

, where $\hat{s}_{noSD}$ refers to the normalized non-SD carrying EEG tracing, $\hat{s}_{SD}$ refers to the normalized SD carrying EEG segment, $\hat{n}$ refers to the normalized random noise and $BPF_{0.5-45Hz}$ refers to the bandpass limiting signal frequency between 0.5 and 45 Hz. By calculating the weighted average across the SD attenuated EEG $\hat{s}_{noSD}|\hat{s}_{SD}|$ and noises $\hat{n}$ through constants $\alpha \in$ $[0,0.3]$ and $\beta \in [0,0.2]$, we then obtain a segment of non-SD EEG undergoing weighted attenuation by a true SD EEG power reduction profile with noises added on through a statistical weighted approach. The rest of the non-SD carrying EEG segments are also mixed and normalized with the noise with its weight constant $\beta \in [0,0.2]$.

By doing so, the generated simulation of SD carrying EEG segments would exhibit different levels of SD signature power suppression and scalp mediated noises through alteration of the weighting constants, which ensuring the non-SD-specific EEG suppression can be inherited in the generated EEG for improved specificity analysis.

To validate the genuinity of the simulated EEG segments, in reference with COSBID consensus guidance [20] in identification of SD in electrophysiology and in line with other common approaches in validating the simulated SD carrying EEG segments [30], [61], we plotted the EEG tracing on time domain and the leaky integral of the EEG power for examination of their features in line with COSBID guidance [20]. The electrophysiological features observed in the EEG wave and leaky integral of EEG power tracing satisfy the COSBID guidance prescribed features for a SD carrying EEG tracings [20], with the clinical electrophysiologists confirming of the validity of the generated SD carrying EEG segments (**Figure 6**).

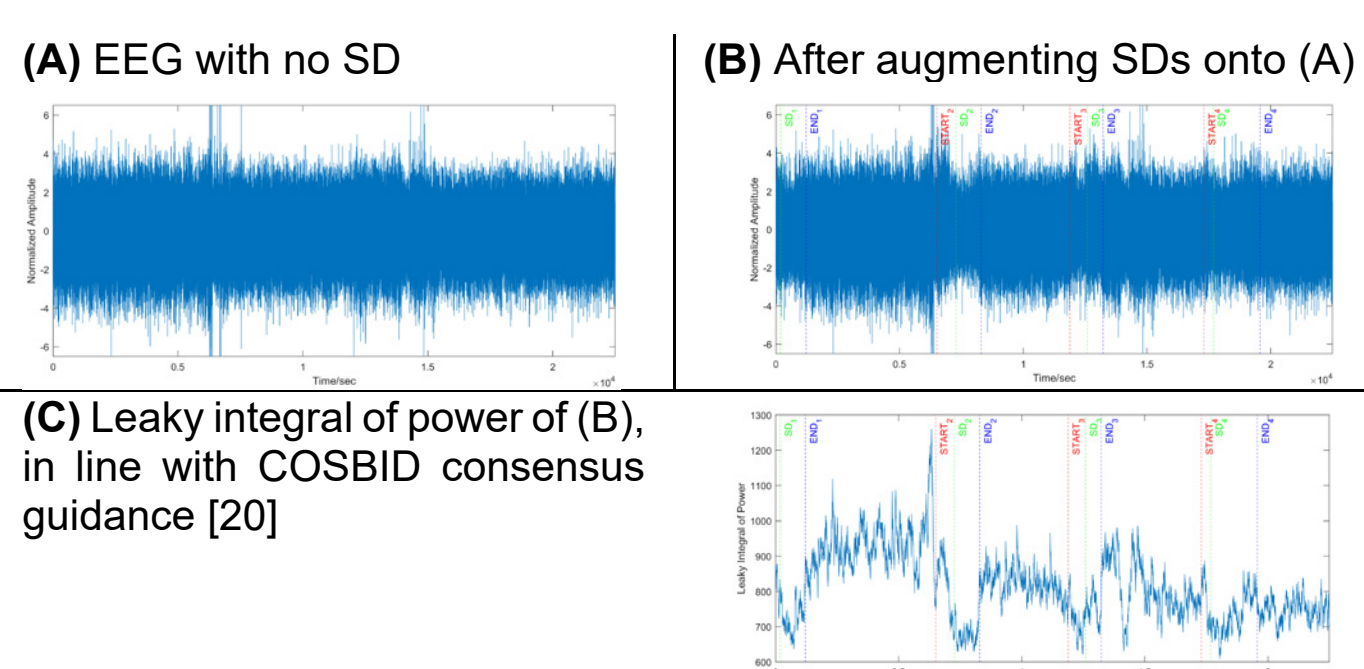

**Figure 6.** A typical example of **(A)** an EEG with no SD and **(B)** EEG augmented with SDs, following the method in IV.A.2). We could see variable AC power offsets in EEGs with no SD **(A)**, where a specific identification of SD characteristic SD depression is the key to ensure SD detection specificity. **(B)** The generated SD carrying EEG simulated the typical clinical scenario where non-SD specific power offset fluctuations mixed up with the SD specific ones, reflecting more realistic observations of EEG over patients experiencing SDs. The time points of the SD start ($START_x$) and end ($END_x$) and the SD depression peak ($SD_x$) are labelled on **(B)**. **(C)** Leaky integral of power of (B), in line with COSBID consensus guidance [20]

### 4) SD detection performance over non-SD EEGs but with brain injury

Apart from testing over the augmented SD carrying EEGs, we also test the model over non-SD carrying EEG tracing, where these augmented EEGs are based on. We evaluated the model over these clinically acquired non-SD carrying EEG (17 hours × 6 channels) with alternating AC power offset to assess the model's ability to differentiate between the SD specific power changes and non-SD neural signal changes, to establish its specificity over negative controls.

### 5) Ablation test: Significance of employing both spectrogram imaging and power vectors for SD detection with high accuracy

To further assess the necessity of inputting both spectrogram imaging and power vectors, we performed training and testing through either only the imaging path (**Figure 3C.1**, inputting just the spectrograms) or only the vector path (**Figure 3C.2**, inputting just the conventionally used power vectors) and computed evaluation metrics for each of them. This will also allow us to assess if the frequency band restricted spectrogram imaging can be used as a standalone biomarker.

## B. Evaluation metrics

We evaluated the performance by 3 key metrics.

### 1) Stage 5 binary outcome evaluation

For stage 5 of the framework **(Figure 3.5)**, where it generates a series of binary detection outcomes as to if there is a SD in the 30-min time window around each time point, we calculated the specificity, sensitivity and accuracy of the generated binary results compared to the ground truth, as per Equations (4-6).

$$\text{Sensitivity} = \text{TP}/(\text{TP+FN}) \quad (4)$$

$$\text{Specificity} = \text{TN}/(\text{TN+FP}) \quad (5)$$

$$\text{Accuracy} = (\text{TP+TN})/(\text{TP+FN+TN+FP}) \quad (6)$$

, where TP=true positive, TN=true negative, FP=false positive and FN=false negative.

### 2) Stage 6 confidence score evaluation

For stage 6 of the framework **(Figure 3.6)**, where it adds up surrounding binary results through the 30-min sliding window as per **III.E**, we calculated the Euclidean distance between the generated series of confidence scores and the expected series of confidence scores. The Euclidean distance $d(\text{u}, \text{v})$ between two 1-D arrays $\text{u}, \text{v}$ can be calculated as per Equation (7).

$$d(\text{u}, \text{v}) = \sqrt{\sum_i |\text{u}_i - \text{v}_i|^2} \quad (7)$$

As discussed earlier, the expected series of confidence scores would have 30 at the peak of SD suppression and linearly descent to 0 15mins before and after the peak point.

### 3) Stage 6 confidence peak visual comparison

We also carried out visual inspection to count the true positive peaks (detection result peaks overlapping with the ground truth peaks), false negative peaks (where no detection result peak associated with a ground truth peak) and false positive peaks (where a standalone detection result peak is found with no ground truth peak associated) (**Figure 7**). With counts of true positives and false negatives, the specificity rate of detection result peaks could also be calculated from there.

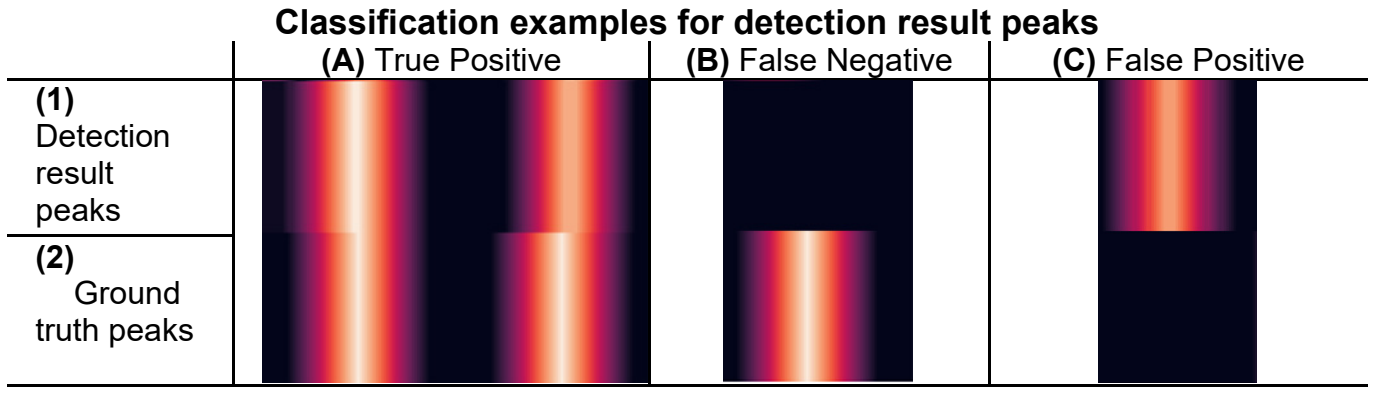

**Figure 7.** Examples of classifications of **(A)** true positive peaks (detection result peaks overlapping with the ground truth peaks), **(B)** false negative peaks (where no detection result peak associated with a ground truth peak) and **(C)** false positive peaks (where a standalone detection result peak is found with no ground truth peak associated).

TABLE I

PERFORMANCE OF THE MAIN MODEL AND OTHER ABLATED STANDALONE MODELS OVER AUGMENTED SD CARRYING EEG

| Model inputs | Stage 5 binary results | | | Stage 6 confidence scores | Stage 6 confidence score peaks | | | |
|---|---|---|---|---|---|---|---|---|
| | Specificity | Sensitivity | Accuracy | Euclidean Distance | No. of True positives | No. of False Negatives | No. of False positives | Sensitivity |
| Spectrogram imaging only | 0.8078 | **0.8325** | 0.8148 | 2364.2 | **949** | **2** | 155 | **0.9978** |
| Power vector only | 0.8559 | 0.6175 | 0.7879 | 2655.7 | 893 | 58 | 185 | 0.9390 |
| Spectrogram image + power vector | **0.9155** | 0.7095 | **0.8567** | **1898.4** | 941 | 10 | **113** | 0.9848 |

TABLE II

PERFORMANCE OF THE MAIN MODEL AND OTHER ABLATED STANDALONE MODELS OVER EEG WITH NO SD

| Model inputs | Stage 5 binary results | Stage 6 confidence scores | Stage 6 confidence score peaks |
|---|---|---|---|
| | Specificity | Euclidean Distance | False positives |
| Spectrogram imaging only | 0.8570 | 605.2 | 53 |
| Power vector only | 0.8850 | 550.7 | 39 |
| Spectrogram image + power vector | **0.9369** | **363.3** | **31** |

TABLE III

PREPROCESSING AND INFERENCE TIME OF THE MODEL FRAMEWORK IF WITH NVIDIA T4 GPU AND IF WITH CPU ONLY OVER EEG INPUT OF 98,900 MINUTES.

| | Total (98,900 mins) | | Average time per hour | |
|---|---|---|---|---|
| Preprocessing | 129 sec | | 0.08 sec | |
| Model inputs | With NVIDIA T4 GPU | | CPU only | |
| | Total (98,900 minutes) | Average per hour | Total (98,900 minutes) | Average per hour |
| Spectrogram imaging only | 5 sec | 3 msec | 190 sec | 0.12 sec |
| Power vector only | 5 sec | 3 msec | 141 sec | 0.086 sec |
| Spectrogram image + power vector | 6 sec | 4 msec | 220 sec | 0.13 sec |

## C. Model Training Strategy

The model is trained by Adam optimizer [62] and the loss of binary cross-entropy [63]. The hyperparameters were set as: learning rate= 0.001, $\beta_1$=0.5, $\beta_2$=0.5. All models are trained on a NVIDIA A100 GPU for 20 epochs with a batch size of 320.

## D. Results

### 1) SD detection performance over augmented SD carrying EEG

The performance of the ablated standalone image path model and the ablated standalone power vector path model over the augmented EEGs with SDs are shown in **Table I**. Visualization of the confidence scores generated are displayed in **Figure 8A**.

It suggested that our model has high specificity (0.9155), sensitivity (0.7095) and accuracy (0.8567) in producing binary outcomes for SD detection in **Stage 5**, although the ablated model with only spectrogram imaging input can offer higher sensitivity (0.8325). Our model offered the lowest Euclidean distance (1898.4) in confidence score generation in **Stage 6**. When viewed from the visualization of confidence score peaks, our model offered similar sensitivity (0.9848) in highlighting SD depression peaks in **Stage 6** to the highest performing ablated model (0.9978) and the fewest false positive peaks.

In **Figure 8A**, we could observe seldom appearance of false positive peaks of moderate confidence scores in our model, while the ablated ones gave much more false positive peaks and more false negatives.

### 2) Model performance in excluding SDs over the non-SD EEG but with brain injury

The performance of our model, the ablated standalone image path model and the ablated standalone power vector path model over the EEG with no SD can be found in **Table II**. Visualization of the confidence scores generated can be found in **Figure 8B**.

It is suggested that our model performed the best with the highest specificity (0.9369) in producing binary outcome series in **Stage 5**, smallest Euclidean distance (363.4), and the fewest false positive peaks (31) in **Stage 6**.

In **Figure 8B**, we could observe seldom appearance of false positive peaks of moderate confidence scores in our model, while the ablated ones gave much more false positive peaks.

### 3) Ultra-Quick Inference Time

To demonstrate the light weight of the model to fulfil the clinical objective of instant processing and real-time monitoring. We re-run the model inference on a GPU of a lower specification, NVIDIA T4 16 GB, which is more widely available, to assess model's performance for more expanded use. A batch size of 30 is selected as the model require a detection outcome array of 30 minutes for Stage 6 confidence score calculations.

To further simulate the real-world clinical setting, we also infer the model over the machine without the use of GPU, where our machine is equipped with a CPU of Intel Xeon CPU of 2.20 GHz. The batch is reduced to 1 to further save CPU memory.

The model was input with 98,900 minutes of EEG, where we took their average time of processing. The results can be found in **Table III**. We could observe that all model cost very little time for processing using a low specification GPU, 3-4 milliseconds per each hour of EEG in average. Even without a GPU, all model's average inference time are below 0.2 second.

Even adding up with the average preprocessing time of 0.08 second per each hour of EEG, the total processing time to process each hour of EEG remain less than 0.3 second, which is extremely fast and almost instant. Such ultra-quick processing time showcased its potential for expanded clinical use and implementation of such algorithms.

# V. DISCUSSION

This study introduces a novel ultra-light-weight dual-path model for non-invasive SD detection from EEG using frequency-restricted 2-D spectrogram images and 1-D power vectors. Our method performance surpasses conventional EEG signal power vectors. Unlike prior approaches [30], [33], [34], our model adopts a single electrode approach, accommodating flexible EEG map positioning without density assumptions or AC power offset. Notably, our efficient model processes in <0.3 sec per EEG hour, with just 30 mins of monitoring required for a decision. This offers broader clinical access for EEG-based SD detection compared to lengthy methods [34].

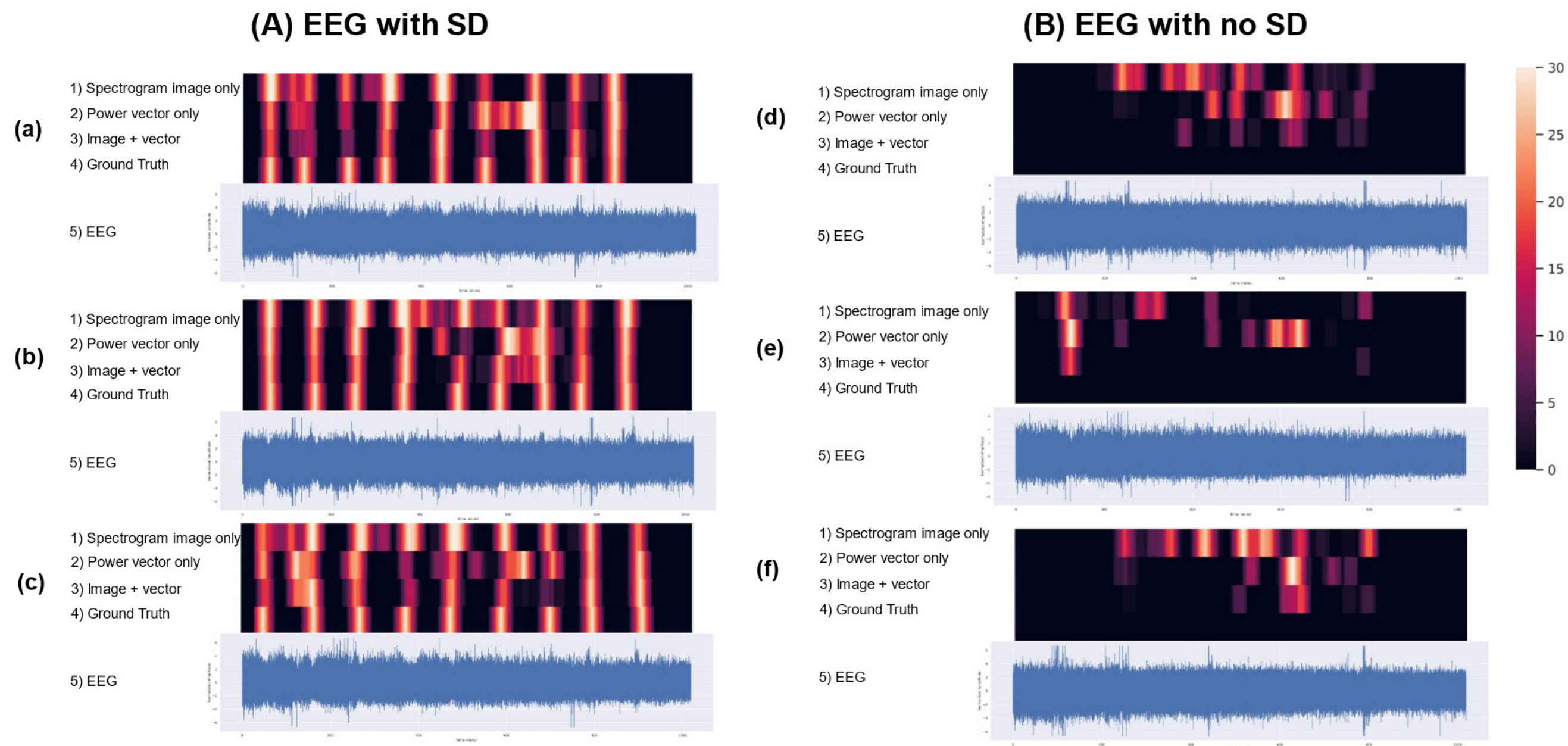

**Figure 8. (a-f)** Visualization of confidence scores of example cases for **(A)** EEG with SD and **(B)** EEG with no SD through ablated models with either **(1)** only spectrogram image input or **(2)** only power vector input, **(3)** our designed model combing both image and vector inputs, compared to the visualization of the **(4)** expected confidence scores, and **(5)** EEG of these example cases.

### *A. Performance Analysis*

This dual-path model has been evaluated on SD carrying EEG and EEG with no SD. We found the method exhibits extremely high specificity and competent sensitivity for EEG based non-invasive SD detection (**Table I, Figure 8A**). On EEG with no SD, the dual-path model showed high specificity, demonstrating its ability to differentiate between SD-specific and non-SD-specific EEG power change (**Table II, Figure 8B**), which is the usual limitation of previous method with only power vector input [33].

The model's lightweight nature extends its use on diverse bedside machines, benefiting more patients. Unlike prior models, ours achieves real-time SD detection without needing high-end hardware like GPUs. It makes decisions in <0.3 sec (average per EEG hour), with just 30 mins of continuous monitoring. This contrasts with the conventional method's 2-hour inference and 30-hour EEG requirement. Our model represents a significant step towards real-time non-invasive SD monitoring.

### *B. Significance of employing both spectrogram imaging and power vectors for model input*

Through comparing our image-vector dual-path model with the two ablated models (each with either the 2-D spectrogram image or the 1-D power vector as the sole input), we demonstrated the superiority of the dual-path model over single-path models for its higher specificity and accuracy and competent sensitivity (0.9848) for SD depression peak visualization and its shortest Euclidean distance to the expected confidence score distribution in Stage 6 (**Table I, Figure 8A**). When assessed over EEG with no SD, the dual-path model exhibited the highest specificity and the smallest confidence scores for false positives, demonstrating its high specificity over SD detection (**Table II, Figure 8B**).

### *C. EEG spectrogram as a new reliable independent biomarker for SD*

This also demonstrated the ability of the spectrogram imaging as a standalone biomarker for SD detection with the highest sensitivity (0.9978) for SD depression peak visualization, although with a much lower specificity. When compared with the conventionally used power vector when it is used as a standalone biomarker, the spectrogram imaging showed a much higher sensitivity and accuracy in SD detection and a lower Euclidean Distance to the expected confidence score distribution (**Table I**). Although the model with spectrogram input alone led to a lower specificity rate when generating Stage 5 binary outcome series, the subsequent summation windowing process made its count of false positive peaks to be lower than the output from the model with power vector alone at Stage 6. With a lower count of false positive peaks and higher sensitivity in confidence peak generation at Stage 6, by this study we can safely suggest the superiority of spectrogram imaging over the conventionally used power vector. This can be seen in the visualization of the confidence scores generated (**Figure 8A**).

Although **Table II** showed the model with standalone spectrogram image input had more false positive peaks and longer Euclidean distance from the expected confidence score distribution, if we calculate the average Euclidean distance per false positive peaks, we found the average Euclidean distance per false positive peak is lower for the model with standalone spectrogram inputs (11.4) than the one with the standalone power vector input (14.1). This illustrated the spectrogram image input model's lower confidence towards its false positive result peaks, making the false positive peaks easier for exclusion upon further thresholding.

### D. Towards balancing specificity and sensitivity of SD detection and illustrating the strength and time duration of each SD: thresholding on the confidence scores

The continuity of the confidence scores generated offer clinicians extra support when assessing the detection outcomes, while giving us another opportunity to re-analyze the results and to re-balance the specificity and sensitivity of SD detection through setting thresholds over the confidence score generated. In **Figure 9**, we can see that, if a typical confidence score series is thresholded, thresholding confidence scores can help to calibrate the position of the SD peak, improving localization of the peak point (**Figure 9:** Red and white arrows). Thresholding can also filter out low-confidence false positive peaks (**Figure 9**: Pink and yellow arrows), which can also be seen when the models were tested over EEGs with no SD (**Figure 10**). However, thresholding of an inappropriately higher level can turn true positive peaks of moderate confidence levels into false negative results (**Figure 9**: Green and blue arrows), where a more careful choice of a fixed threshold level upon further expanded studies is recommended.

Beyond binary SD detection, thresholding can reveal SD duration (or width) for each episode. Where more adaptive thresholding methods should be used to define SD width considering score trend and strength. Confidence scores can also relate to AC power depression and SD strength, pending validation on larger datasets.

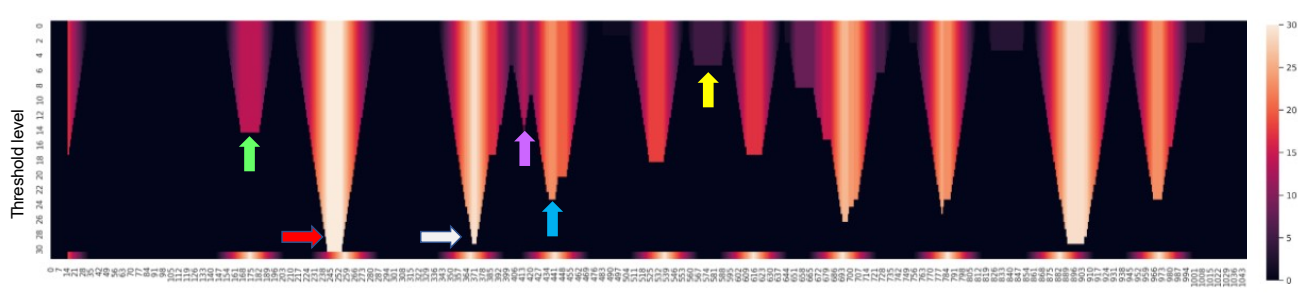

**Figure 9.** Visualization of confidence scores thresholded at different levels for potential improvement of specificity and sensitivity. **(a)** Red and white arrows: Thresholding confidence scores can help to calibrate the position of the SD peak, improving localization of the peak point. **(b)** Pink and yellow arrows: Thresholding can also filter out low-confidence false positive peaks. **(c)** Green and blue arrows: However, thresholding of an inappropriately higher level can turn true positive peaks of moderate confidence levels to false negative results.

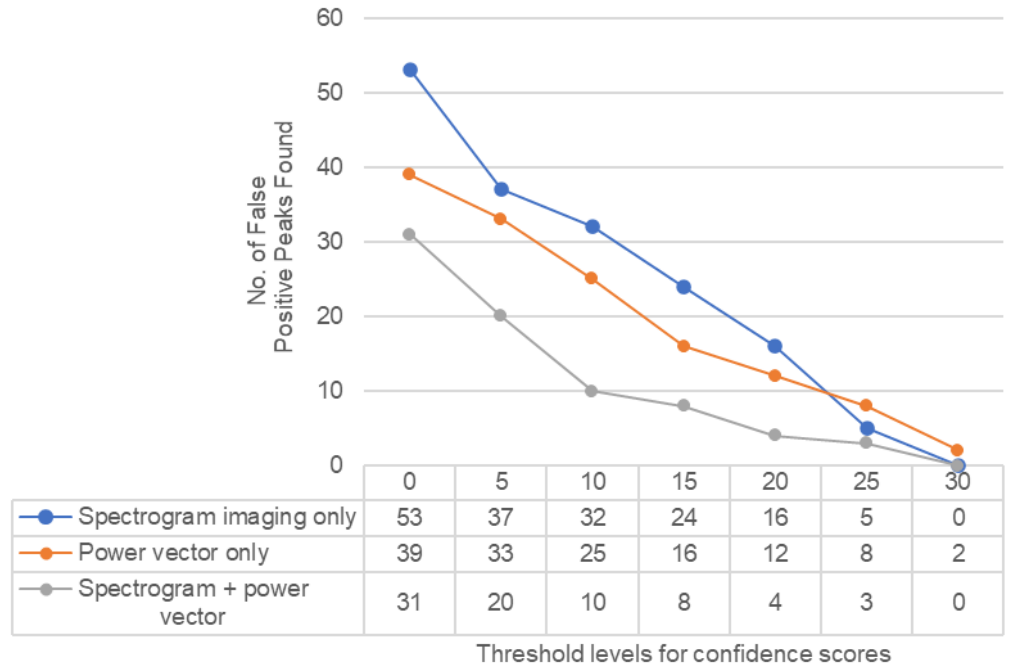

| | 0 | 5 | 10 | 15 | 20 | 25 | 30 |
|---|---|---|---|---|---|---|---|
| Spectrogram imaging only | 53 | 37 | 32 | 24 | 16 | 5 | 0 |
| Power vector only | 39 | 33 | 25 | 16 | 12 | 8 | 2 |
| Spectrogram + power vector | 31 | 20 | 10 | 8 | 4 | 3 | 0 |

**Figure 10.** Number of false positive result peaks (Stage 6) if thresholded with different levels

### E. Clinical contribution, limitation, and future work

This model provides a computationally light-weight processing method for real-time detection of SDs over EEG, which has demonstrated its efficacy for SD detection.

In the clinical context of neurocritical monitoring, real-time SD monitoring and instant brain injury prognosis is the key. The ultra-short processing time of <0.3 sec per each hour of EEG in average and the light weight of the model framework makes it the first real-time non-invasive SD detection framework for expanded clinical use and translation.

Our single electrode approach also provides early evidence that multiple electrodes may not be necessary for SD detections. As part of the proposal, we introduced a new biomarker – spectrogram for additional dimension for consideration and processing, which outperforms the conventionally used 1D series power vectors in SD detections per our results. We would then recommend the use of spectrogram to extract extra SD specific features over the feature axis.

We showed that, with a non-standard dense map of EEG electrodes, SDs should be present in recorded EEG signals in a clearly defined way. Furthermore, the proposed detection model showed that SDs can be clearly identified specifically. Therefore, in principle the model should be easily transferable for screening other patients. For the model's further generalizability and transferability, the following extensions will need to be made – **(1)** having patient populations with more diverse injury types (TBI, MHS, and aSAH cases); **(2)** obtaining data from different vendors (i.e., different kinds of electrodes from different manufacturers and different amplifiers); **(3)** evaluating our method with multiple clinical centers (to account for variations in clinical practice). This would then need to be repeated in a larger number of patients to build a compelling clinical case for the use of our methods.

In this study, we calculated the evaluation metrics for all SDs without further classifying SD subtypes. We anticipate further study will calculate their evaluation metrics for each type of SD (isolated, continuous and intermediate [25]) to facilitate a more extensive understanding of the model's prediction outcome.

We recognise that the craniectomy (i.e. removal of the bone flap before surgical closure) of the patients may have facilitated the cortical potentials reaching the scalp electrodes and thus discovery of these SD signature events in EEG. However, the breach effects associated with the craniectomy for amplification of EEG signals is of concern more for higher frequency activities such as in alpha frequency band (8-12 Hz) [64], [65], whereas we only used the frequency band between 0.5 and 1.85 Hz as described in III.B, so we do consider the aforementioned concern unlikely. In contrast, the craniotomy (where the bone flap **is** replaced) will most likely leave a substantial volume of air beneath the bone flap, insulating scalp electrodes from brain electrophysiology, which may bring challenges to read EEG from these patients.

It should also be noted that in this study the EEG electrodes were placed on the sculp with the underlying skull removed during the data acquisition, which might facilitate the EEG signals as well and enhancing the results. However, as there is abundant evidence suggesting that the key EEG suppression pattern can still be observed in EEG even though the skull is intact [29] or bone flap is restored [24], [25], we are confident that the model will work for patients with intact or restored skull, despite of possible extra training and fine tuning.

In this study, we only had patients with low GCS scores enrolled. We are also keen to understand the effects of SD for patients with high GCS and for patients who are not in drug induced coma, as their EEG distribution across different frequency bands will be markedly different. By having higher GCS scores, voluntary movements may affect the EEG tracings, which we may need to re-train the model or even adopt an even more complex model for SD identifications.

Moreover, more data from patient with no SDs may be needed, for further justification of the specificity of the model for clinical use.

## VI. Conclusion

This study presents a novel image-vector dual-path model that simultaneously processes 2-D EEG spectrogram imaging and 1-D EEG temporal power vector for non-invasive SD detection from EEG. A unique approach transforms 1-D EEG into 2-D spectrogram, boosting detection efficacy with an added frequency axis. Unlike prior methods [30], [33], [34], our technique doesn't constrain EEG mapping, density, or assume AC power offset. Our lightweight model is a first step for real-time monitoring and instant SD detection, benefiting broader clinical use. We highlight frequency-restricted spectrogram imaging's superiority as a standalone biomarker compared to conventional EEG signal power vectors. Recommending its use for EEG-based SD detection, it adds a frequency dimension to SD observation.